\documentstyle[here,epsfig]{mn2e}
\def\simlt{\mathrel{\rlap{\lower 3pt\hbox{$\sim$}}\raise 2.0pt\hbox{$<$}}}
\def\simgt{\mathrel{\rlap{\lower 3pt\hbox{$\sim$}} \raise 2.0pt\hbox{$>$}}}

\def\gtsima{$\; \buildrel > \over \sim \;$}
\def\ltsima{$\; \buildrel < \over \sim \;$}
\def\gtrsim{\lower.5ex\hbox{\gtsima}}
\def\lesssim{\lower.5ex\hbox{\ltsima}}
\def\url#1{{\ttfamily\def\/{/\discretionary{}{}{}}#1}}
\def\etal{et al.~}

\def\eg{{\frenchspacing e.g. }}

\begin{document}

\newcommand{\q}{\begin{equation}}
\newcommand{\qa}{\begin{eqnarray}}
\newcommand{\qs}{\begin{eqnarray*}}
\newcommand{\nq}{\end{equation}}
\newcommand{\nqa}{\end{eqnarray}}
\newcommand{\nqs}{\end{eqnarray*}}
\newcommand{\ud}{\mathrm{d}}

\title[DM decays/annihilations and first structures] 
{The impact of dark matter decays and annihilations on the formation of the first structures}
\author[E. Ripamonti, M. Mapelli, A. Ferrara]
{E. Ripamonti$^{1}$, M. Mapelli$^{2}$,  A. Ferrara$^{2}$\\
$^{1}$Kapteyn Astronomical Institute, University of Groningen, Postbus 800, 9700 AV, Groningen, The Netherlands; {\tt ripa@astro.rug.nl}\\
$^{2}$SISSA, International School for Advanced Studies, Via Beirut 4, 34100, Trieste, Italy\\}

\maketitle \vspace {7cm }

\begin{abstract}
We derive the effects of dark matter (DM) decays and annihilations on
structure formation. We consider moderately massive DM particles
(sterile neutrinos and light DM), as they are expected to give the
maximum contribution to heating and reionization.  The energy injection
from DM decays and annihilations produces both an enhancement in the
abundance of coolants (H$_2$ and HD) and an increase of gas temperature.
We find that for all the considered DM models the critical halo mass for
collapse, $m_{crit}$, is generally higher than in the unperturbed
case. However, the variation of $m_{crit}$ is small. In the most extreme
cases, i.e. considering light DM annihilations (decays) and halos
virializing at redshift $z_{vir}>30$ ($z_{vir}\sim{}10$), $m_{crit}$
increases by a factor $\sim{}$4 ($\sim{}$2). In the case of
annihilations the variations of $m_{crit}$ are also sensitive to the
assumed profile of the DM halo.  Furthermore, we note that the fraction
of gas which is retained inside the halo can be substantially
reduced (to $\approx 40$ per cent of the cosmic value), especially in
the smallest halos, as a consequence of the energy injection by DM
decays and annihilations.
\end{abstract}

\section{Introduction}\label{introduction section}
One of the fundamental questions concerning the formation of first
structures is the minimum halo mass (critical mass, $m_{crit}$) for
collapse at a given redshift (Silk 1977; Rees \& Ostriker 1977;
White \& Rees 1978; Couchman 1985; Couchman \& Rees 1986; de Araujo \&
Opher 1988, 1991; Haiman, Thoul \& Loeb 1996).

Tegmark et al. (1997; T97) thoroughly addressed such question,
pointing out how $m_{crit}$ crucially depends on the abundance of H$_2$,
the main coolant present in the metal free Universe. Subsequent studies
(Abel et al. 1998; Fuller \& Couchman 2000; Galli \&{} Palla 1998, 2002;
Ripamonti 2006) refined the model of T97, accounting also for minor
effects, such as the cooling induced by HD molecules.

The production of molecules and $m_{crit}$ are sensitive to any physical
process which can release energy in the intergalactic medium (IGM). In
fact, the injection of energy in the IGM can  either delay the
collapse of first halos (because of the increased gas temperature, or of
photodissociation of molecules) or favour structure formation (because of the
enhancement in the abundance of free electrons, which act as catalysts
for the formation of molecules).

For this reason, it is crucial to understand the influence
of reionization sources on structure formation. Many studies have shown
that massive metal free stars are efficient in dissociating H$_2$
molecules, quenching star formation in the first halos (Haiman, Rees \&
Loeb 1997; Ciardi, Ferrara \& Abel 2000; Ciardi et al. 2000; Haiman,
Abel \& Rees 2000; Ricotti, Gnedin \& Shull 2002; Yoshida et
al. 2003). Intermediate mass black holes, produced by the collapse of
first stars, are thought to efficiently re-heat the IGM, increasing
$m_{crit}$ and reducing star formation in the smaller mass halos
(Ricotti \& Ostriker 2004; Ricotti, Ostriker \& Gnedin 2005; Zaroubi et
al. 2006).

Also particle decays and annihilations can be sources of
partial reionization and heating (see Mapelli, Ferrara \& Pierpaoli 2006
and references therein), and could influence structure formation.
For example, Shchekinov \& Vasiliev (2004) investigated the possible
effect on $m_{crit}$ due to ultra-high energy cosmic rays (UHECRs)
emitted by particles decaying in the early Universe. Biermann \& Kusenko
(2006) considered the impact on structure formation due to sterile
neutrino decays. Both these studies found a substantial enhancement on
the abundance of molecular coolants (H$_2$ and/or HD).
However, they neglected the possible increase of gas temperature due to
UHECRs or decays, respectively. 

More recently, Stasielak, Biermann \&
Kusenko (2006) evaluated the effect of sterile neutrino decays
accounting also for the heating of the gas. However, their single-zone
model  is likely to oversimplify the crucial behaviour of gas
density during the halo collapse.

In this paper, we consider the influence of dark matter (DM) decays and
annihilations on structure formation, taking into account variations
induced both in the chemical and in the thermal evolution of the IGM and
of the gas inside halos. Furthermore, we substitute the single-zone
models, which are commonly adopted in previous papers (Haiman et
al. 1996 is an important exception), with more sophisticated 1-D
simulations. We focus on relatively low mass DM particles, such as
sterile neutrinos and light DM (LDM), as their effect on the IGM is
expected to be much more important than that of heavier
($\gtrsim{}100$ MeV) DM particles (Mapelli et al. 2006).

Sterile neutrinos are expected to decay into active neutrinos and
keV-photons (Dolgov 2002), while LDM can either decay or annihilate
producing electron-positron pairs (Boehm et al. 2004; Hooper \& Wang
2004; Picciotto \& Pospelov 2005; Ascasibar et al. 2006). keV-photons
interact with the IGM both via Compton scattering and photo-ionization;
instead, the electron-positron pairs undergo inverse Compton scattering,
collisional ionizations, and positron annihilations (Zdziarski \&
Svensson 1989; Chen \& Kamionkowski 2004; Ripamonti, Mapelli \& Ferrara
2006, hereafter RMF06). RMF06 derived the fraction $f_{abs}(z)$ of
energy emitted by sterile neutrino decays and LDM decays or
annihilations which is effectively absorbed by the IGM through these
processes. In this paper, we adopt the fits of $f_{abs}(z)$ given by
RMF06.

In Section 2 we describe the hydro-dynamical code used to derive
$m_{crit}$ and the DM models which we adopt. In Section 3 we discuss the
effect of DM decays and annihilations on the chemical and thermal
evolution of the IGM, giving an estimate of the Jeans mass. In Section 4
we describe the chemical and thermal evolution of the gas inside the
halos, deriving $m_{crit}$. In the discussion (Section 5) we address
various points, such as the variations in the baryonic mass fraction
inside the halos induced by DM decays/annihilations and the influence of
the concentration of the DM profile.

We adopt the best-fit cosmological parameters after the 3-yr WMAP
results (Spergel et al. 2006), i.e. $\Omega{}_{\rm b}=0.042$,
$\Omega{}_{\rm M}=0.24$, $\Omega{}_{\rm DM}\equiv{}\Omega{}_{\rm
M}-\Omega{}_{\rm b}=0.198$, $\Omega{}_\Lambda{}=0.76$, $h=0.73$,
$H_0=100\,{}h$ km s$^{-1}$ Mpc$^{-1}$.

\section{Method}\label{section method}
\subsection{The code}

In order to estimate the effects of the energy injection from DM decays
and annihilations, it is necessary to follow the chemical and thermal
evolution of primordial gas. This can be done with single-zone models
such as the one originally described in T97, or the adaptations by
Shchekinov \& Vasiliev (2004), and by Stasielak et al. (2006). However,
this kind of models is bound to use some approximations which can be
very crude. First of all, single-zone models cannot follow the density
evolution of a virialized halo because of their lack of "resolution", so
that it is usually assumed that after virialization the gas density in a
halo is both {\it uniform} in space and {\it constant} in time. Even in
the linear phase of the collapse, the top-hat model provides a
reasonable description of the DM component, but it becomes a rough
approximation when the baryonic component is considered, as
hydro-dynamical effects are likely to become important at scales below
the Jeans or the filtering length (Peebles 1993; Gnedin 2000).

An exact description would require a  3-D simulation (such
as those of Abel, Bryan \& Norman 2002, and of Bromm, Coppi \& Larson
2002); but it is possible to capture the basic features of the collapse
phenomenon by an intermediate, time-efficient approach, e.g. by means of
1-D simulation, as they are both more accurate than single-zone models,
and much faster than  3-D simulations (Haiman et al. 1996).

In this paper, we use the 1-D Lagrangian, spherically symmetric code
described by Ripamonti et al. (2002), as updated in Ripamonti
(2006). Such a code includes the treatment of:
\begin{itemize}
\item{}{the gravitational and hydro-dynamical evolution of the gas (by
means of an artificial viscosity scheme);}
\item{}{the chemical evolution of 12 species (H, H$^+$, H$^-$, H$_2$,
H$_2^+$, D, D$^+$, HD, He, He$^+$, He$^{++}$, and $e^-$; see table 1 of
Ripamonti 2006 for a list of the considered reactions and of the adopted
reaction rates);}
\item{}{the cooling (or heating) due to a number of components, such as
the compressional (adiabatic) heating, the emission and absorption of
line radiation from H, H$_2$ and HD (accounting for the effects of the
cosmic microwave background, CMB), the heating (or cooling) from the
Compton scattering of CMB photons off free electrons, and the heating
(or cooling) due to chemical reactions (e.g. the formation or
dissociation of H$_2$ molecules);}
\item{}{the gravitational effects of DM, according to a simple model
which is based on the top-hat formalism up to the turn-around redshift,
smoothly evolving into a concentrated profile after virialization (see
Section \ref{DM_profiles_section} for more details).
}
\end{itemize}

\subsection{DM profiles}
\label{DM_profiles_section}
The code does not include a self-consistent treatment of DM. Instead,
the function $\rho_{\rm DM}(r,z)$, describing the DM density profile and
its redshift evolution, must be chosen {\it a priori}. Since our results
might depend on this choice, we decided to study two quite different
cases, which we call `isothermal' and `NFW' (from the profile
described in Navarro, Frenk \& White 1997) depending on the shape of the
density profile after virialization.

In both cases the DM distribution was assumed to be spherically
symmetric, and concentric with the simulated region, whose central part
represents the halo which is being investigated. At any redshift a DM
mass $M_{\rm DM}=M_{\rm halo}\Omega_{\rm DM}/\Omega_{\rm M}$ is assumed
to be within the truncation radius
\begin{equation}
  R_{\rm tr}(z) = 
  \left\{{
    \begin{array}{ll}
      \left({{3\over{4\pi}}
	{{M_{\rm DM}}\over{\rho_{\rm TH}(z)}}}\right)^{1/3}
      & {\rm if\ }z \geq z_{\rm ta}\\
      R_{\rm vir}
      \left[{2-{{t(z)}\over{t(z_{\rm vir})-t(z_{\rm ta})}}}\right]
      & {\rm if}\ z_{\rm ta} > z \ge z_{\rm vir}\\
      R_{\rm vir}
      & {\rm if}\ z < z_{\rm vir}\\
  \end{array}}\right .
\end{equation}
where $t(z)$ is the time corresponding to redshift $z$, and $z_{\rm
vir}$, $z_{\rm ta}$, $R_{\rm vir}$ and $\rho{}_{\rm TH}(z)$ are the halo
virialization and turn-around redshifts, its virial radius, and the DM
density inside the halo at $z>z_{\rm ta}$ (as derived from the evolution
of a simple {\it top-hat} fluctuation; see e.g. Padmanabhan 1993, or
T97), respectively. The exact definition of these quantities can be found
in Ripamonti 2006.

The isothermal and NFW assumptions differ only
after the halo virialization ($z<z_{\rm vir}$); both of them refer to a
static DM profile. In the isothermal case
\begin{equation}
\rho{}_{\rm DM}(r,z) =
      	\left\{{
	\begin{array}{ll} 
	\rho{}_{\rm core}
	    & {\rm if}\ r \leq R_{\rm core};\\
	\rho{}_{\rm core} (r/R_{\rm core})^{-2}
	    & {\rm if}\ R_{\rm core} \leq r \leq R_{\rm tr};\\
	\rho_0\Omega_{\rm DM}(1+z)^3
	    & {\rm if}\ r > R_{\rm tr};\\
        \end{array}}\right .
\label{DM_density_profile_iso}
\end{equation}
instead, in the NFW case the DM density profile is chosen to be
\begin{equation}
\rho_{\rm DM}(r,z) =
      \left\{{
      \begin{array}{ll} 
      {\rho_{\rm NFW}\over{(r/R_{\rm core})(1+r/R_{\rm core})^2}}
           & {\rm if}\ r \leq R_{\rm tr};\\
	\rho_0\Omega_{\rm DM}(1+z)^3
	    & {\rm if}\ r > R_{\rm tr};\\
        \end{array}}\right .
\label{DM_density_profile_NFW}
\end{equation}
where $\rho_0\simeq 1.88\times10^{-29}\,h^2\;{\rm g\,cm^{-3}}$ is the
critical density of the Universe at present.  In both cases $R_{\rm
core}=\xi R_{\rm vir}$, where $\xi{}=0.1$ is a parameter (cfr. Hernquist 1993, and Burkert 1995 for the choice of its value; in
the NFW case $\xi$ the inverse of the more commonly uses concentration
parameter), and the densities $\rho_{\rm core}$ and $\rho_{\rm NFW}$
can be found by requiring the DM mass within $R_{\rm tr}$ to be equal to
$M_{\rm DM}$.

At $z>z_{\rm vir}$ both the isothermal and NFW case assume the DM density
profile described by equation~\ref{DM_density_profile_iso}. However, at
such redshift the profile is not static, because the core radius is
evolved with redshift
\begin{equation}
  R_{\rm core}(z) = 
  \left\{{
    \begin{array}{ll} 
      R_{\rm tr}(z) & {\rm if\ } z \ge z_{\rm ta}\\
      R_{\rm vir}
      \left[{2-
	       {{\left({2-\xi}\right)\,{}t(z)}\over{t(z_{\rm vir})-t(z_{\rm ta})}}}\right]
      & {\rm if\ } z_{\rm ta} > z \ge z_{\rm vir}\\
  \end{array}}\right .
\label{rcore_definition}
\end{equation}
Such a choice combines the behaviour of a {\it top-hat} fluctuation (the
density inside $R_{tr}$ is assumed to be uniform until the turn-around)
with a transition to the final density profiles. 

We only considered the case $\xi=0.1$ (i.e. concentration 10 for the NFW
profile), instead of varying $\xi{}$, because the differences between the isothermal and NFW
cases are quite relevant even with the same value of $\xi$. In fact,
the NFW case is representative of concentrated halos, whereas the
isothermal case is representative of relatively shallow potentials.

However, it is important to note that the flat central profile of the
isothermal case helps to ensure that the behaviour we observe near the
centre is due to the self-gravity and hydrodynamics of the {\it
simulated} gas, rather than to the {\it assumed} DM profile.

\subsection{Treatment of the DM energy injection}

The above code was modified in order to include the effects of the energy
injection from DM decays/annihilations on both the chemical and
thermal evolution of the gas. 

The gas can be heated, excited and ionized by the energy input due to DM
decays/annihilations. It is important to note that the fraction of the
absorbed energy going into each one of these components is quite
unrelated to how the energy was deposited in the IGM at the first
step. For example, if a keV a photon ionizes an atom, the resulting
electron will generate a cascade of collisions, and the energy of the
photon will go not only into ionizations, but also into excitations and
heating.

Thus, given the energy injection per baryon from DM decays and
annihilations, $\epsilon(z)$ (described in Section
\ref{DM_models_section}), we split it into an heating and ionization
component\footnote{The sum of these two components is less than
$\epsilon(z)$, as a significant fraction of the injected energy goes in
atomic/molecular excitations and does not affect the chemical or thermal
state of the gas.}

\begin{eqnarray}
\epsilon_{heat}(z) & = &
\tilde{C}[1-(1-x(z))^{\tilde{a}}]^{\tilde{b}} \epsilon(z)\\
\epsilon_{ion}(z) & = & {{1-x(z)}\over{3}} \epsilon(z),
\end{eqnarray}
where $x(z)$ is the ionization fraction. In the first equation we are
using the fit to the results of Shull \& Van Steenberg (1985) which is
provided in their paper (with $\tilde{C}=0.9971$, $\tilde{a}=0.2663$,
and $\tilde{b}=1.3163$), while in the second equation we are using the
fit to the same results given by Chen \& Kamionkowski (2004).

The heating component is simply added to the equations describing the
thermal state of the gas. Instead, the ionization
component is further split between H, He, He$^+$, D, H$_2$, HD and
H$_2^+$, according to their number abundance:
\begin{equation}
\epsilon_i(z) = {{\tilde{N}_i}\over{
    \sum_{j\in(H,He,He^+,D,H_2,HD,H_2^+)} \tilde{N}_j}} \epsilon_{ion}(z)
\end{equation}
where the indices $i$ and $j$ indicate chemical species, and
$\tilde{N}_i \equiv N_i$ if the species $i$ is atomic, or as
$\tilde{N}_i \equiv 2 N_i$ if the species $i$ is molecular; $N_i$ is the
number density (per unit volume) of the chemical species $i$. In
principle, the terms in the sum above should be weighted by the
cross-section for each species. However both the cross-section and the
energy spectrum are too complex to be implemented in our simple
calculations. In particular, the energy spectrum is expected to be the
result of a cascade (Shull \& Van Steenberg 1985). The error in
neglecting these factors is quite small, as $m_{crit}$ is more sensitive
to the temperature increase than to the chemistry (see Section 4).

\begin{table}\label{tab_1}
\begin{center}
\caption{List of chemical reactions stimulated by the DM energy injection.
}
\begin{tabular}{lll}
\hline
Species & Reaction & Threshold \\
\hline
H       & H $\rightarrow{}$ H$^+$ + $e^-$ & 13.6 eV\\
He      & He $\rightarrow{}$ He$^+$ + $e^-$ & 24.6 eV\\
He$^+$  & He$^+$ $\rightarrow{}$ He$^{++}$ + $e^-$ & 54.4 eV\\
D       & D $\rightarrow{}$ D$^+$ + $e^-$ & 13.6 eV\\
H$_2$   & H$_2$ $\rightarrow{}$ H + H & 4.48 eV\\
HD      & HD $\rightarrow{}$ H + D & 4.51 eV\\
H$_2^+$ & H$_2^+$ $\rightarrow{}$ H + H$^+$ & 2.65 eV\\
\hline
\end{tabular}
\end{center}
\end{table}

The quantity $\epsilon_i$ approximates the energy which is absorbed by
chemical reactions dissociating the species $i$, and is translated
into a reaction rate (number of reactions per particle per unit time)
through a division by the energy threshold $E_{th,i}$ of the considered
reaction. The list of the reactions and of the energy thresholds is
given in Table~1.

We neglect the absorption of ionization energy by H$^+$, He$^{++}$,
$D^+$ and $e^-$, because these species cannot be ionized further. We
also neglect the ionization energy absorption by H$^-$ because the
energy threshold for the transformation of H$^-$ into H is negative, and
cannot be treated with our simple formalism. However the number
abundance of H$^-$ is always very small, and the number of dissociations
induced by DM decays/annihilations is likely to be negligible.

\subsection{DM models}
\label{DM_models_section}
We apply this formalism to two different DM candidates, i.e. sterile
neutrinos and LDM. Sterile neutrinos are one of the most popular warm DM
 (WDM) candidates (Colombi, Dodelson \& Widrow 1996; Sommer-Larsen \& Dolgov
2001; Bode, Ostriker \& Turok 2001). They can decay via different channels (Dolgov 2002; Hansen \&
Haiman 2004). In this paper we are interested in the radiative decay,
i.e. the decay of a sterile neutrino into a photon and an active
neutrino, because of its direct impact on the IGM (Mapelli \& Ferrara
2005; Mapelli et al. 2006). The photon produced in the decay interacts
with the IGM both via Compton scattering and photo-ionization (RMF06).

LDM particles have recently become of interest, because they provide a
viable explanation for the detected 511-keV excess from the Galactic
centre (Boehm et al. 2004; Kn\"odlseder et al. 2005). If they are source
of the 511-keV excess, then their maximum allowed mass $m_{\rm LDM}$
should be 20 MeV, not to overproduce detectable gamma rays via internal
Bremsstrahlung (Beacom, Bell \& Bertone 2004). If we consider also the
production of gamma rays for inflight annihilation of the positrons,
this upper limit might become $\sim{}$3 MeV (Beacom \& Y\"uksel 2006).

LDM can either decay or annihilate, producing photons, neutrinos and
pairs.  In this paper we consider both LDM decays and annihilations, but
we restrict their treatment to the case where the product particles are
$e^{+}-\,{}e^{-}$ pairs. In fact, in the case of pair production the
impact of LDM on the IGM is maximum (RMF06). The $e^{+}-\,{}e^{-}$ pairs
are expected to interact with the IGM via inverse Compton scattering,
collisional ionization and positron annihilation (RMF06).

\begin{table*}\label{tab_2}
\begin{center}
\caption{ List of the main characteristics of the considered DM models. In particular, from the leftmost to the rightmost column: DM model, $n_{{\rm DM},0}$, $\tau{}_{\rm DM}$ (if decaying particle), $\langle{}\sigma{}\,{}v\rangle{}$ (if annihilating),
 comoving free-streaming lengths ($\lambda_{\rm FS,n}$ and $\lambda_{\rm FS,i}$) and the associated
  mass scale. $\lambda_{\rm FS}$ is taken to be the maximum between
  $\lambda_{\rm FS,n}$ and $\lambda_{\rm FS,i}$.}
\begin{tabular}{lcccccc}
\hline
DM model & $n_{{\rm DM},0}/10^3$ & $\tau{}_{\rm DM}/(10^{27}\textrm{ s})$ & $\langle{}\sigma{}\,{}v\rangle{}/(10^{-29}\textrm{ cm}^3\textrm{ s}^{-1})$ & $\lambda_{\rm FS,n}$(pc) & $\lambda_{\rm FS,i}$(pc) &
$M_{\rm FS}(M_\odot)$ \\
\hline
$\nu$ 4 keV (decaying)    & 1176  & 2.96  & $-$ & $3\times{}10^5$ & $60$ & $5\times10^8$ \\
$\nu$ 15 keV (decaying)   & 314   & 1.98  & $-$ & $8\times{}10^4$  & $35$ & $1\times10^7$ \\
$\nu$ 25 keV (decaying)   & 188   & 0.097 & $-$ & $4.8\times{}10^4$  & $25$ & $2\times10^6$ \\
LDM 3 MeV (decaying)      & 1.49  & 1.2   & $-$ & 2.2 & $<0.1$ & $2\times10^{-7}$ \\
LDM 10 MeV (decaying)     & 0.446 & 4.0   & $-$ & 0.5 & $<0.1$ & $3\times10^{-8}$ \\
LDM 1 MeV (annihilating)  & 4.46  & $-$   & 4   & 9.5 & 300 & 0.5 \\
LDM 3 MeV (annihilating)  & 1.49  & $-$   & 12  & 2.2 & 100 & 0.02 \\
LDM 10 MeV (annihilating) & 0.446 & $-$   & 24  & 0.5 & 30 &  5$\times10^{-4}$ \\
\hline
\end{tabular}
\end{center}
\end{table*}

\subsubsection{The energy input from the ``background''}

We first consider the energy injected in the general IGM after cosmic DM
decays/annihilations.

Both in the case of sterile neutrinos and of LDM, the rate of energy
transfer (per baryon) to the IGM because of this ``background''
contribution can be written as:
\begin{equation}
\epsilon{}_{bkg}(z)=f_{abs}(z)\,{}\dot{n}_{\rm DM}(z)\,{}m_{\rm DM}\,{}c^2,
\end{equation}
where $m_{\rm DM}$ is the mass of a DM particle and $c$ is the speed of
light. The energy absorbed fraction, $f_{abs}(z)$, has been derived in
RMF06; $\dot{n}_{{\rm DM}}(z)$ is the decrease rate of the number of DM
particles per baryon.

In the case of DM decays, $\dot{n}_{{\rm DM}}(z)$ is given by
\begin{equation}\label{DM density decay}
\dot{n}_{\rm DM}(z) \simeq {{n_{{\rm
	DM},0}}\over{\tau_{\rm DM}}},
\end{equation}
where $n_{{\rm DM},0}$ and $\tau_{\rm DM}$ are the current number of DM
particles per baryon and the lifetime of DM particles,
respectively. $\tau_{\rm DM}$ is assumed to be much longer than the
present value of the Hubble time, as is the case for all the models we
are considering.

For the annihilations: 
\begin{equation}\label{DM density annih}
\dot{n}_{\rm DM}(z) \simeq {1\over2}\, n_{{\rm DM},0}^2\,{}N_{\rm b}(0)\,{}
\langle \sigma\,v\rangle (1+z)^3,
\end{equation}
where $N_{b}(0)=2.5\times{}10^{-7}\;{\rm g\,cm^{-3}}$ is the current
baryon number density (Spergel et al. 2006), and $\langle\sigma v\rangle$ is the
thermally averaged DM annihilation cross-section.

Both for sterile neutrinos and LDM, the values of $n_{{\rm DM},0}$,
$\tau_{\rm DM}$ and $\langle\sigma v\rangle$ adopted in this paper are
the same reported in RMF06; for convenience, they are listed in Table~2.

\subsubsection{The ``local'' energy input}

In addition to the background energy injection discussed above, the
baryons inside a halo absorb extra energy from the additional
decays/annihilations of overdense halo DM. In the case of decays the
total excess energy ``produced'' inside the halo is
\begin{eqnarray}
E_{loc}(z) & = & {{4\pi\, N_b(0)(1+z)^3\, m_{{\rm DM}}\,
    c^2}\over\tau_{\rm DM}}\nonumber\\
& \times & \int_0^{R_{tr}(z)} dr\,
r^2 [n_{\rm DM}(r,z)-n_{{\rm DM},0}],
\end{eqnarray} 
and in the case of annihilations
\begin{eqnarray}
E_{loc}(z) & = & 4\pi\, [N_b(0)(1+z)^3]^2\, m_{{\rm DM}}\, c^2\,
\langle \sigma v \rangle \nonumber\\
& \times & \int_0^{R_{tr}(z)} dr\,
\frac{1}{2}\,r^2\, [n_{{\rm DM}}^2(r,z)-n_{{\rm DM},0}^2(z)],
\end{eqnarray} 
where $n_{\rm DM}(r,z)=\rho_{\rm DM}(r,z)/[m_{\rm DM} N_b(0)(1+z)^3]$ is
the number of DM particles per baryon at redshift $z$ and at a distance
$r$ from the centre of the halo.

Then, we compute the baryon column density $\Sigma_b$ from the halo centre
to the truncation radius, and find the fraction of the energy
$E_{loc}$ which is absorbed by such a column density, $f_{loc}$. If the DM
produces photons of energy $E_\gamma$,
\begin{equation}
f_{loc} = \Sigma_b \left[{\sigma_{He+H}(E_\gamma) +
\sigma_T {{E_\gamma}\over{m_e c^2}}
g\left({{{E_\gamma}\over{m_e c^2}}}\right)} \right]
\end{equation}
where $\sigma_{He+H}(E)$ is the photo-ionization cross section (see
Zdziarski \& Svensson 1989; RMF06), $\sigma_T$ is the Thomson cross
section, and the function $g$ is defined in equation 4.9 of Zdziarski \&
Svensson (1989).  Instead, if the DM produces electron-positron pairs
with Lorentz factor $\gamma$ (and energy E=$\gamma\, m_e\, c^2$),
\begin{equation}
f_{loc} = \frac{\Sigma_b}{c\,{}N_b(0)(1+z)^3}\,{}\phi_{e,ion}(z,E)+
{{R_{tr}(z)}\over c} \phi_{e,com}(z,E)
\end{equation}
where $\phi_{e,ion}(z,E)$ and $\phi_{e,com}(z,E)$ give the fraction of
the energy of an electron/positron which is absorbed per unit time by
the IGM, because of collisional ionizations and of Compton scattering of
CMB photons, respectively. Such functions are given by equations 14-15,
and 16-18 of RMF06.

We then assume that all the baryons inside the halo absorb the same
amount of energy from this local contribution. So, the ``local'' energy
deposition in each baryon within the truncation radius $R_{tr}(z)$ is
\begin{equation}
\epsilon_{loc}(z,r) = E_{loc} f_{loc} {{M_{gas}[R_{\rm tr}(z)]}\over{m_H}}
\end{equation}
where $M_{gas}(R_{\rm tr})$ is the mass of the gas inside the truncation
radius, and $m_H$ is the mass of an H atom. Instead, for baryons at a
distance larger than $R_{tr}(z)$ from the centre of the halo we assume
$\epsilon_{loc}(z,r)=0$.

The total energy input per baryon from DM decays/annihilations is then
\begin{equation}
\epsilon(r,z) = \epsilon_{bkg}(z) + \epsilon_{loc}(r,z).
\end{equation}

\subsection{The free-streaming/damping lengths}

We need to take into account that small DM fluctuations might be washed
out by damping mechanisms, and in particular by free-streaming (see \eg
Padmanabhan 1993; Sommer-Larsen \& Dolgov 2001; Boehm \etal 2005).

The free-streaming scale depends on the considered particle, and on the
strength of its interactions. If such interactions are negligible, the
comoving free-streaming length is
\begin{equation}
  \lambda_{\rm FS,n} \simeq
  \left\{{
    \begin{array}{ll}
     0.3 \left({{m_{\rm DM}}c^2\over{4\,{\rm keV}}}\right)^{-1}
      \left({\langle{}p\,{}c/(k_B\,{}T)\rangle{}\over{3.15}}\right)\,{} {\textrm{\footnotesize{Mpc}}}\quad
      {\textrm{\footnotesize{for neutrinos}}}\\
      0.20\, (\Omega_{\rm DM} h^2)^{1/3}
      \left({{m_{\rm DM}c^2}\over{1\,{\rm keV}}}\right)^{-4/3} {\textrm{\footnotesize{Mpc}}}\quad
      {\textrm{\footnotesize{for LDM,}}}\\
  \end{array}}\right .
\label{l_free_streaming_std}
\end{equation}
where $m_{\rm DM}$ is the mass of the considered DM particle, $p$ and $T$ are the modulus of the proper momentum and the temperature of the particle, respectively.
The above expressions are derived from Abazajian, Fuller \& Patel (2001), and from Boehm et al. (2005), respectively.

However, if the DM particle interacts at a non negligible rate, the
comoving free-streaming length is (Boehm \etal 2005; Boehm \& Schaeffer 2005)
\begin{equation}
\lambda_{\rm FS,i} = 0.3
\left({{m_{\rm DM}c^2}\over{1\,{\rm MeV}}}\right)^{-1/2}
\left({\tilde\Gamma_{\rm dec,DM}}\over
     {6\times10^{-24}\,{\rm s^{-1}}}\right)^{1/2}
{\rm Mpc}
\label{l_free_streaming_inter}
\end{equation}
with
\begin{equation}
\tilde\Gamma_{\rm dec,DM} = 
\Gamma_{\rm DM}(z_{\rm dec})(1+z_{\rm dec})^{-3},
\end{equation}
where $\Gamma_{\rm DM}(z_{\rm dec})$ is the DM interaction rate at its
decoupling redshift $z_{\rm dec}$.

In the case of decaying particles, we assume that this rate is simply
the inverse of the present lifetimes of DM particles; so, $\Gamma_{\rm
DM}(z_{\rm dec}) < 10^{-24}\,{\rm s^{-1}}$ in all the cases we are
considering (cfr. Table 2). Since $z_{\rm dec} \gg 10^3$,
equations (\ref{l_free_streaming_std}) and (\ref{l_free_streaming_inter})
imply that $\lambda_{\rm FS,i} \ll \lambda_{\rm FS,n}$; so, for decaying
particles we will use a free-streaming scale
$\lambda_{\rm FS}=\lambda_{\rm FS,n}$.

Instead, when annihilating particles are considered, the interaction
rate is $\Gamma_{\rm DM}(z_{\rm dec}) = {1\over2} n_{\rm DM,0} N_{\rm
b}(0) \langle{}\sigma{}v \rangle{}_{\rm dec} (1+z_{\rm dec})^3$, where
$\langle \sigma v \rangle{}_{\rm dec}\sim 10^{-26} \; {\rm
cm^3\,s^{-1}}$ (see, for example, Ascasibar et al. 2006) is the
thermally averaged annihilation cross section at the epoch of
decoupling\footnote{For annihilating particles, a third scale length
might be considered, i.e. the mixed damping length $\lambda{}_{md}$
(Boehm et al. 2003). The upper limit of $\lambda{}_{md}$ is 6.7, 2.9 and
1.2 kpc for 1, 3 and 10 MeV LDM, respectively. These values are a factor
$\sim{}20$ larger than the corresponding $\lambda_{\rm FS,i}$; but our
conclusions about $m_{crit}$ (see Section 4) are not significantly
affected, because $M_{\rm FS}$ is still smaller than the mass range we
are considering.}. Then, in the case of annihilating particles,
$\lambda_{\rm FS,i} \gg \lambda_{\rm FS,n}$, and we must assume
$\lambda_{\rm FS}=\lambda_{\rm FS,i}$. In Table~2 we give the detailed
list of the free-streaming lengths for each DM model. The same table
also lists the free-streaming mass scale
\begin{equation}
M_{\rm FS} =
{{4\pi}\over{3}}\left({\lambda_{\rm FS}\over2}\right)^3\rho_0\Omega_{\rm M} h^2.
\end{equation}

Damping erases fluctuations on scales smaller than $\lambda_{\rm
FS}$. Thus, objects of mass $\lesssim M_{\rm FS}$ are unlikely to
form, unless the small scale power spectrum is regenerated by non-linear
effects at low redshift (see Boehm \etal 2005; however, this
regeneration appears to take plate at $z\sim2$, which is much later than
the epoch we are considering).  To introduce
corrections for this effect in our models is beyond the purposes of this
paper.

 Damping might also affect the density profile of halos, erasing cusps on
 scales $\lesssim\lambda_{\rm FS}$. The values of $\lambda_{\rm FS}$ and
 $M_{\rm FS}$ listed in Table~2 show that this effect of damping can be
 important only for sterile neutrinos.
 In the case of isothermal density profiles, we account for it adopting
 the following correction. If the core radius $R_{\rm core}(z)$ (equation~\ref{rcore_definition}) is
 smaller than $\lambda_{\rm FS}/2$, we increase its value to $R_{\rm
core,new}(z)=min(R_{\rm tr}(z),\lambda_{\rm FS}/2$.  Instead, in
 the case of simulations with NFW density profile we do not introduce
 this correction, because these simulations are intended to explore the
 effects of high concentration (see Section 4.2).

\subsection{The simulations}
Our code was used to run a large number of simulations, in order to
explore a wide range of the $z_{vir}-M_{halo}$ parameter space: we
considered halo masses in the range $10^4-10^7\,M_\odot$, and
virialization redshifts between $10$ and $100$.

We actually simulate a mass which is 1000 times higher than that of the
collapsing halo, in order to include in the simulation a mass which is
larger than the cosmological filtering mass (Gnedin 2000); otherwise,
our treatment of hydrodynamics might be incorrect. The simulated object
is divided in 150 shells, whose spacing was chosen so that (i) the
mass of the shells smoothly increases when moving outwards, (ii) the
central shell always encloses a gas mass of $\sim 0.3\,M_\odot$, and
(iii) the central 100 shells initially enclose a mass $\sim M_{halo}$
(including the DM).

The simulations are started at $z=1000$, when we assume that the gas
density is uniformly equal to the cosmological value (whereas the DM
density profile is not  perfectly uniform; see Section 2.2), the gas
temperature is equal to the CMB temperature ($\simeq 2.728\,{\rm
K}$), and adopt the chemical abundances listed in table 2 of Ripamonti
2006\footnote{The most important abundances listed there are the H 
ionization fraction (at $z=1000$)
$N_{H^+}/N_H = 0.0672$, the helium abundance $N_{He}/N_{H}=0.0833$, and
the deuterium abundance $N_D/N_H=2.5\times10^{-5}$.}.

They are stopped either when the gas density reaches the threshold
$\rho/m_H=10^5\,{\rm cm^{-3}}$, or after a time $2 t(z_{vir})$ has
elapsed.

Each set of simulations was repeated for each different DM
decay/annihilation model, and also for the ``standard'' case without any
energy injection from DM, which is used as a reference against which we
compare our results.

\begin{figure}
\center{{
\epsfig{figure=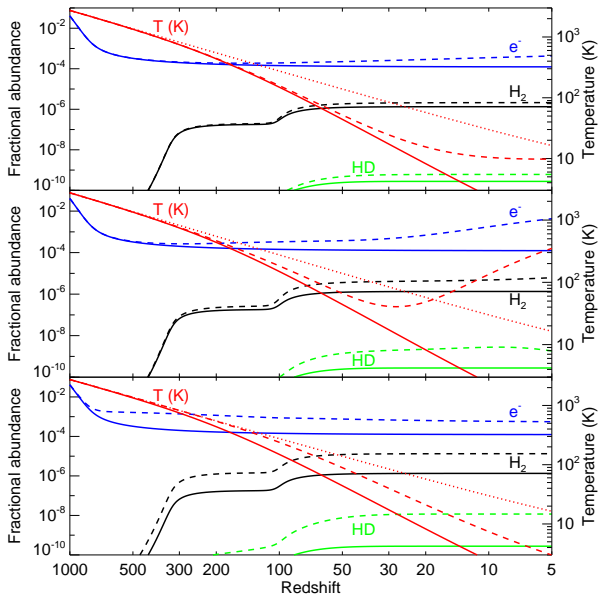,height=8cm}
}}
\caption{\label{fig:fig1}  Effects of decaying/annihilating DM on the IGM evolution. Left axis: fractional abundances of free electrons ($e^-$), H$_2$ and HD as a function of  redshift. Right axis: matter temperature as a function of  redshift. Top panel: Effect of decaying sterile neutrinos of mass 25 keV (dashed line).  Central panel:  decaying LDM of mass 10 MeV (dashed line).  Bottom panel: annihilating LDM of mass 1 MeV (dashed line). In all the panels the dotted line is the CMB temperature and the solid line represents the thermal and chemical evolution without DM decays/annihilations.}
\end{figure}
\section{IGM evolution}
In order to look at the influence of DM decays and annihilations upon the IGM, we have used a simplified version of our code (where the density was assumed to evolve as the cosmological value).
\subsection{Chemistry and temperature}
Our code follows the chemical evolution of 12 chemical species (see previous Section). Two of them, molecular hydrogen (H$_2$) and HD are particularly important for our purposes, because they are the main coolant of the metal free gas.
In Fig.~\ref{fig:fig1} we show the fractional abundances of both H$_2$ and HD together with the ionized fraction and the matter temperature as a function of redshift for the considered DM models.

In all the DM models, both the matter temperature and the abundance of
H$_2$, HD and free electrons are enhanced by DM
decays/annihilations. This effect is smaller for sterile neutrinos than
for LDM particles. The main difference between decays and annihilations
is represented by the redshift range in which the influence of DM is
important. For all the considered quantities (i.e. temperature and
abundance of $e^{-}$, H$_2$ and HD) the energy injection from DM decays
starts to be significant at redshift lower than $\sim{}100$. Instead,
the influence of annihilations is important already at redshift
$\sim{}900$. The annihilations represent also the case where the
abundance of the two coolants is most enhanced (a factor $\sim{}17$ for
the H$_2$ and $\sim{}90$ for the HD). Furthermore, the annihilations
keep the matter temperature close to the CMB temperature everywhere, up
to $z\sim50-100$. This fact can have important consequences for
experiments searching for high redshift HI 21-cm line signals
(Shchekinov \& Vasiliev 2006).

\subsection{Jeans mass}

In order to establish the influence of DM decays or annihilations on the
structure formation, the key point is the following. DM decays and
annihilations increase both the matter temperature and the abundance of
coolants. The former effect tends to delay the formation of structures,
while the latter favours an early collapse of the halos. Which of these
two opposite effects is dominant? When looking at the average properties
of the IGM, the most popular diagnostic is the cosmological Jeans mass,
$m_J$ (Peebles 1993):
\begin{eqnarray}\label{Jeans}
m_J(T,\rho{},\mu{}) & = & \frac{\pi{}}{6}\,{}\left(\frac{\pi{}k_BT}{G\mu{}m_H}\right)^{3/2}\rho^{-1/2}\nonumber{}\\
&\simeq{} & 50\,{}M_\odot{}T^{3/2}\mu{}^{-3/2}\left(\frac{\rho{}}{m_H}\right)^{-1/2},
\end{eqnarray}
where $k_B$ is the Boltzmann constant, $G$ the gravitational constant and $\mu{}$ the mean molecular weight.

\begin{figure}
\center{{
\epsfig{figure=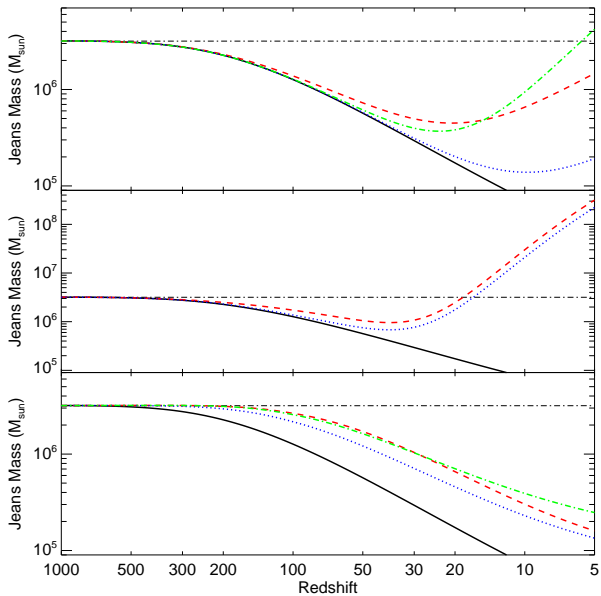,height=8cm}
}}
\caption{\label{fig:figure_jeans}
Jeans mass as a function of redshift. Top panel: Effect of decaying sterile neutrinos of mass 25 (dashed line), 15 (dotted) and 4 keV (dot-dashed). Central panel: Decaying LDM of mass 10 (dashed line) and 3 MeV (dotted). Bottom panel: Annihilating LDM of mass 10 (dashed line), 3 (dotted) and 1 MeV (dot-dashed). In all the panels the solid line represents the thermal and chemical evolution without DM decays/annihilations. The horizontal dot-dashed line shows the behaviour of the Jeans mass calculated by assuming that the gas temperature is always equal to the CMB temperature.
}
\end{figure}
In Fig.~\ref{fig:figure_jeans} we show the evolution of $m_J$.
For all the considered DM models $m_J$ is considerably increased by the
effect of decays and annihilations.  This means that, when the IGM is
considered, the increase of the matter temperature dominates over the
enhancement of the coolant abundance. For example, at $z=10$ $m_J=
3.0\times{}10^7{}M_\odot{}$ in the case of 10-MeV LDM decays, a factor
$\sim{}420$ higher than in the unperturbed case (for which
$m_J=7.2\times{}10^4\,{}M_\odot{}$). The increase of $m_J$ is less
pronounced, but nevertheless significant in the case of sterile neutrino
decays ($m_J=6.6\times{}10^5\,{}M_\odot{}$ for 25-keV sterile neutrinos)
and LDM annihilations ($m_J=3.9\times{}10^5\,{}M_\odot{}$ for 1-MeV
LDM).

From this fact one could naively infer that  DM decays and annihilations strongly delay the formation of the first structures. However, $m_J$ refers only to the global properties of the IGM, and does not account for the non-linear evolution of collapsing halos.

\section{Evolution inside halos}
\begin{figure}
\center{{
\epsfig{figure=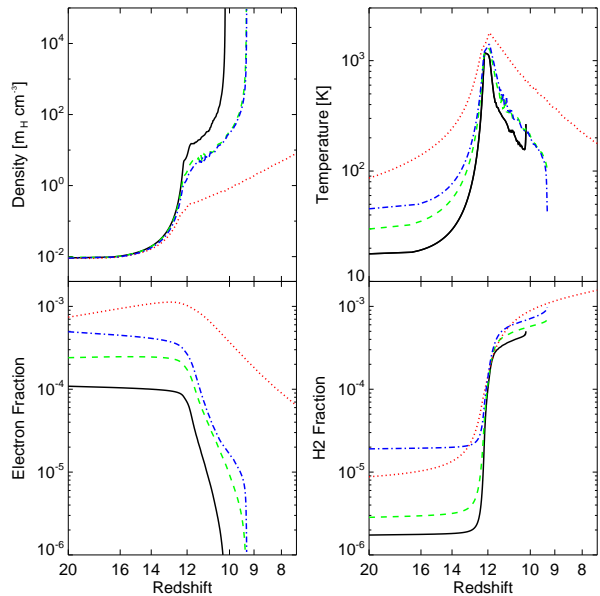,height=8.cm}
}}
\caption{\label{fig:fig3} Evolution of the central region of a
$6\times{}10^5\,{}M_\odot{}$ isothermal halo virializing at
$z_{vir}=12$.  From left to right and from top to bottom: density,
temperature, electron abundance and H$_2$ abundance as function of
redshift. The solid line represents the unperturbed case (i.e. without
DM decays and annihilations). The dashed, dot-dashed and dotted lines
account for the contribution of 25-keV sterile neutrino decays, 1-MeV
LDM annihilations and 10-MeV LDM decays, respectively.  }
\end{figure}
\begin{figure}
\center{{
\epsfig{figure=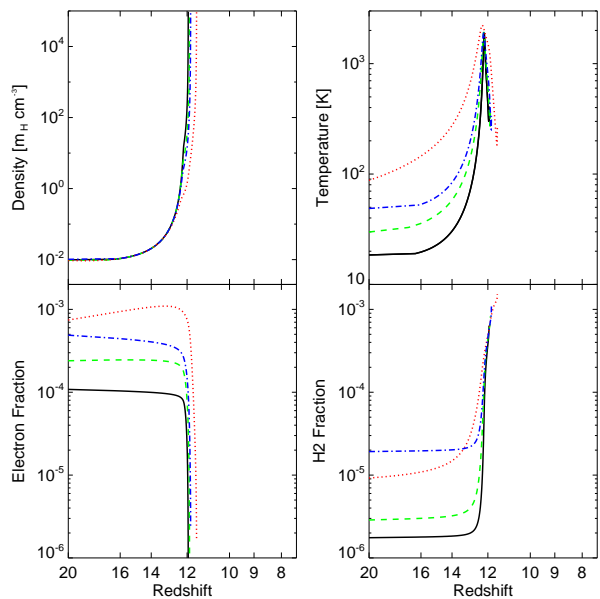,height=8cm}
}}
\caption{\label{fig:fig4} Evolution of the central region of a
$2\times{}10^6\,{}M_\odot{}$ isothermal halo virializing at
$z_{vir}=12$.  From left to right and from top to bottom: density,
temperature, electron abundance and H$_2$ abundance as function of
redshift. The dashed, dot-dashed and dotted lines are the same as in
Fig.~\ref{fig:fig3}.  }
\end{figure}
\subsection{Chemistry and temperature}
To assess the star forming ability of the first halos, it is necessary
to follow their hydro-dynamical and chemical evolution. This has been
done by using the code described in Section~2. Fig.~\ref{fig:fig3}
(Fig.~\ref{fig:fig4}) shows, as an example, the evolution of the most
relevant properties of the gas at the centre of a
$6\times{}10^5\,{}M_\odot{}$ ($2\times{}10^6\,{}M_\odot{}$) halo,
virializing at $z_{vir}=12$. In these figures the effects of different
DM decay/annihilation scenarios are compared with the unperturbed
case. For the smaller halo (Fig.~\ref{fig:fig3}) LDM annihilations and
sterile neutrino decays delay the collapse by $\sim 1$ redshift
units, and in the case of LDM decays this effect is even more
significant: in presence of 10 MeV-LDM decays the halo has not collapsed
yet after one Hubble time from virialization.
Instead, for the larger halo (Fig.~\ref{fig:fig4}) the difference between the case with and without DM decays/annihilations is negligible.

It is worth to discuss in detail the gas temperature ($T$) and number
density ($N=\rho/m_H$) evolution shown in Fig.~\ref{fig:fig3}. At the
epoch of turn-around ($z\sim{}19-20$) all the models have the same $N$;
but in the unperturbed one $T$ is 1.5-5 times lower than in the
others. This is important, because the fast increase in $T$ and $N$ due
to the virialization process is essentially adiabatic, as it happens on
a time-scale much faster than that for cooling.  In the adiabatic
approximation,
$\Gamma{}_{ad}=-P\,{}dV\propto{}P\,{}[N^{-1}-(N+\Delta{}N)^{-1}]$ (where
$P$ is the gas pressure), is proportional to the initial $T$ for any
given increase in $N$. $T$ and $P$ increase faster in models
"pre-heated'' by DM decays/annihilations, so that the pressure gradient
slowing and halting the collapse develops earlier. In fact, the phase of
unimpeded collapse stops at $N\sim{}10$ cm$^{-3}$ for the unperturbed
case, but only at $N\sim{}0.3-3$ cm$^{-3}$ for the other models. At
these densities the cooling per unit mass is proportional both to the
H$_2$ fraction and to $N$, so that the higher central density of the
model without DM decays/annihilations largely compensates its lower
H$_2$ abundance, i.e. the unperturbed case is the {\it fastest} cooling
one. However, it is interesting to note that the final $T$ is much lower
for the cases where DM energy input is included, because of the
enhancement of the HD fraction, which provides an efficient cooling
mechanism also at $T\lesssim{}200$ K (Ripamonti 2006).


\subsection{Critical mass}
Each of our simulated halos was classified as collapsing (or,
equivalently, efficiently cooling) or non-collapsing (inefficiently
cooling), depending on whether it reaches a maximum density larger than
10$^5\,{}m_H$ cm$^{-3}$ in less than a Hubble time. We define the
critical mass, $m_{crit}$, as the minimum mass of a collapsing halo at a
given virialization redshift $z_{vir}$.



\begin{figure}
\center{{
\epsfig{figure=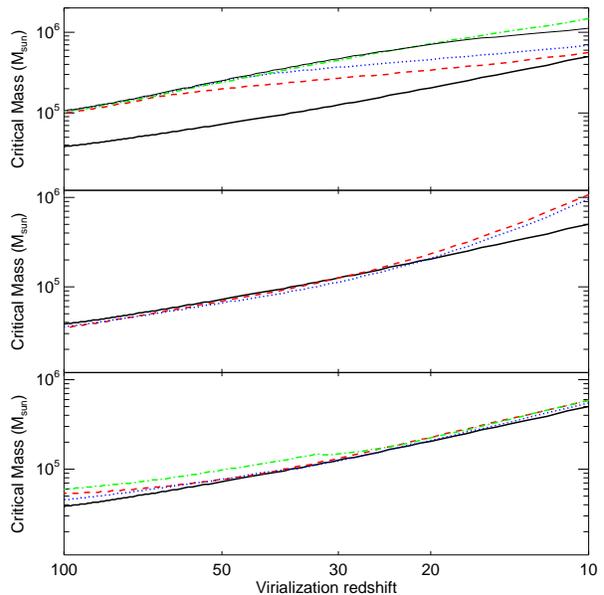,height=8cm}
}}
\caption{\label{fig:fig5} The critical mass as a function of the
virialization redshift $z_{vir}$ for isothermal halos. Top panel: Effect
of decaying sterile neutrinos. Central: Decaying LDM. Bottom:
Annihilating LDM. The lines used in the three panels are the same as in
Fig.~2. The thin solid line used in the top panel represents the
case of non-decaying WDM (with mass 4 keV).}
\end{figure}
\begin{figure}
\center{{
\epsfig{figure=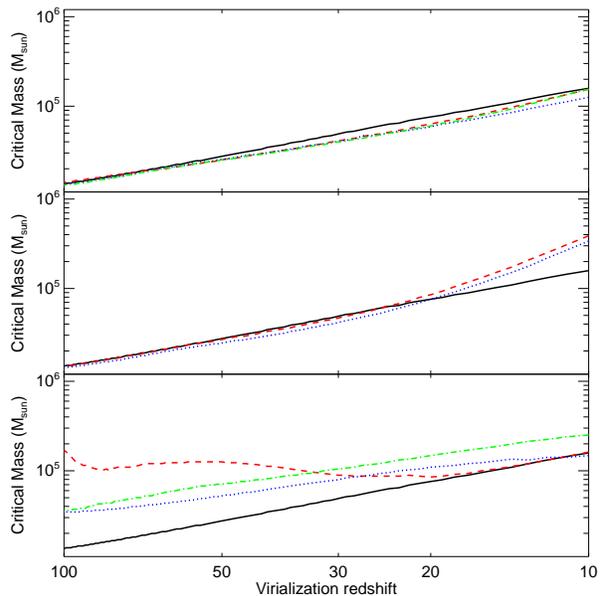,height=8cm}
}}
\caption{\label{fig:fig5bis} The critical mass as a function of the
virialization redshift $z_{vir}$ for NFW halos. The lines used in the
three panels are the same as in Fig.~\ref{fig:fig5}. 
}
\end{figure}
The values of $m_{crit}$ as a function of the virialization redshift are
shown in Fig.~\ref{fig:fig5} for isothermal halos, and in
Fig.~\ref{fig:fig5bis} for NFW halos. The main trend inferred from the
behaviour of $m_J$ (i.e. the delay of structure formation) is confirmed
by $m_{crit}$. However, the influence of DM decays and annihilations on
$m_{crit}$ is much smaller than  could be
expected from $m_J$.

In the case of isothermal halos, the effects of the DM decays on
$m_{crit}$ (top and central panel) are completely negligible at high
redshift and become significant only at redshift less than 20. Even for
$z_{vir}=10$ and 10-MeV decaying LDM (which has the strongest effect),
$m_{crit}$ increases only of by a factor 2.

Sterile neutrinos (especially with mass of 4 keV) seem to significantly
increase $m_{crit}$. However, this is not connected with sterile
neutrinos decays; but it is mainly due to differences in the assumed DM density
profile, as the free-streaming length for these WDM halos is larger
than the ``standard'' value of the core radius ($0.1R_{\rm vir}$) and
the ``damping'' correction described in Section 2.5 is important. 
In
fact, $m_{crit}$ is almost the same down to $z\sim{}20$, if we consider (thick dot-dashed line in
Fig.~\ref{fig:fig5}) or neglect (thin solid line) sterile neutrino
decays, and the difference between these two cases is small even at lower redshift.

In the case of LDM annihilations in isothermal halos, $m_{crit}$
is higher than in the unperturbed case for every considered
virialization redshift ($z_{vir}=10-100$), confirming that the
annihilations play a role even at very high $z$. However, the difference
with respect to the unperturbed $m_{crit}$ is always less than a factor
$\sim{}2$.

If we assume a NFW profile for DM halos (Fig.~\ref{fig:fig5bis}), the
general effect is a substantial decrease of $m_{crit}$ (independently of
DM decays/annihilations), due to the higher central densities, and the
consequent stronger gravitational pull. LDM decays tend to increase
$m_{crit}$ (at $z<20$) with respect to the unperturbed case; the
increase factor is only marginally larger than in the case of isothermal
halos. Sterile neutrino decays appear to slightly reduce $m_{crit}$ with
respect to the unperturbed case; but this result is unphysical, as it
does not account for damping. It is worth to note that, even if we
ignore damping effects and adopt the models with  the highest  possible
concentration, the influence of sterile neutrino decays on structure
formation remains very small.

Instead, in the case of LDM annihilations, $m_{crit}$ generally
increases more in NFW halos than in isothermal halos. This is due to the
importance of the local contribution for annihilations (see Section
5.2).

\begin{figure}
\center{{
\epsfig{figure=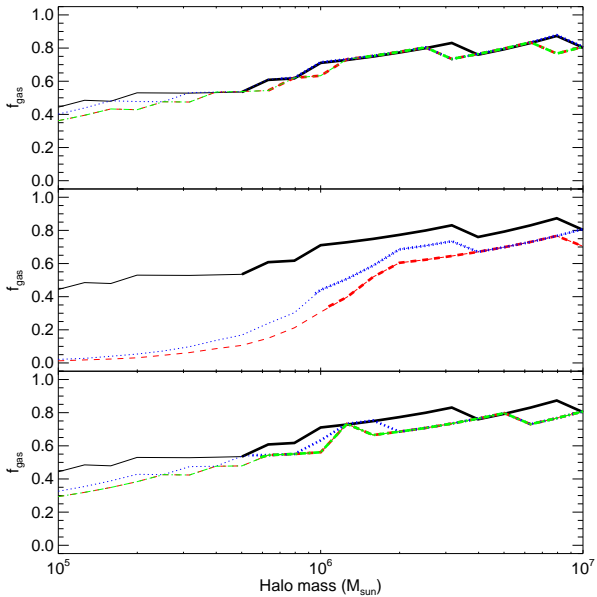,height=8cm}
}}
\caption{\label{fig:fig6} Halo baryonic mass fraction (see definition in
equation~\ref{eq:fraction_gas}) as a function of the halo mass for a
fixed virialization redshift ($z_{vir}=10$) and for an isothermal
DM profile. Top panel: Effect of decaying sterile neutrinos. Central:
Decaying LDM. Bottom: Annihilating LDM. The lines used in the three
panels are the same as in Fig. 2. Thick (thin) lines indicate that the
halo mass is larger (smaller) than $m_{crit}$.  }
\end{figure}

\section{Discussion}
\subsection{Gas fraction}
Why considering either $m_J$ or $m_{crit}$ leads to such widely
different conclusions?  As mentioned earlier, $m_J$ is a rough tracer of
the minimum halo mass for collapse; but it is not sensitive to the local
properties of the halo. For example, to calculate $m_J$ one has to
assume that the density is uniform.  On the contrary, $m_{crit}$ depends
on the evolution of the central region of the halo, where the density
increases much more rapidly than in the outskirts, driving the collapse
(Figs.~\ref{fig:fig3} and \ref{fig:fig4}).  When the density becomes
sufficiently high, the molecular cooling largely overcomes the heating
due to DM decays/annihilations (Figs.~\ref{fig:fig3} and \ref{fig:fig4}).
This is the reason why $m_{crit}$ does not increase significantly in 
presence of DM decays/annihilations. Instead, $m_J$, which does not
account for the density increase and for the consequent cooling
enhancement, dramatically grows.
Then, we can conclude that $m_J$ does not provide a realistic estimate of
the effects of DM decays/annihilations.

Is there any physical implication of the growth of $m_J$?
$m_J$ is directly related to the global hydro-dynamical
behaviour of the gas inside a halo: in halos with mass below $m_J$
the gas pressure prevents the development of a large
gas overdensity, while in more massive halos the gas accumulation
should proceed almost unimpeded. If so, the mass fraction in baryons
within the virial radius of a halo should be of the order of the
cosmological average $\Omega_b/\Omega_M$ when $M_{halo}\gtrsim m_J$,
but it should be lower when $M_{halo}\lesssim m_J$. To check this hypothesis, we derive the baryonic mass fraction ($f_{gas}$) normalized to $\Omega{}_b/\Omega{}_M$, i.e.
\begin{equation}\label{eq:fraction_gas}
f_{gas}=\frac{M_{gas}(R_{vir})}{M_{{\rm
halo}}}\,{}\frac{\Omega{}_{M}}{\Omega{}_b},
\end{equation}

where $M_{gas}(R_{vir})$ is the mass of gas within the virial radius at
the final stage of each simulation; instead $M_{halo}$ is the total mass of the halo at the beginning of the simulation.

In Fig.~\ref{fig:fig6} we show $f_{gas}$ as a
function of mass inside halos virializing at redshift $z_{vir}=10$. The
comparison between the various scenarios clearly shows that the energy
injection from DM decays/annihilations can substantially reduce the gas
fraction inside all halos, especially the smallest ones. This is clearly
related to the increase in $m_J$, as the largest variation in $f_{gas}$
occurs in the LDM decay scenario, where the increase of $m_J$ is maximum
(Fig.~\ref{fig:figure_jeans}).

\begin{figure}
\center{{
\epsfig{figure=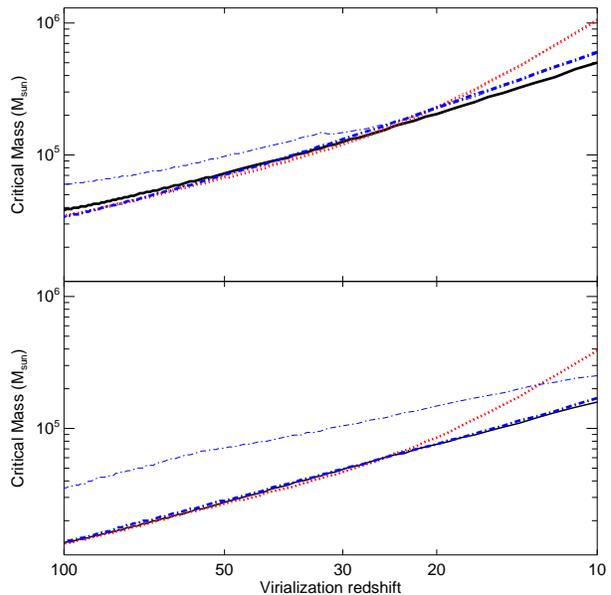,height=8.cm}
}}
\caption{\label{fig:fig7} Evolution of $m_{crit}$ as a function of the
virialization redshift in the case without DM (solid lines),
with 3-MeV LDM decays (dotted) and with 1-MeV LDM annihilations
(dot-dashed). Thin (thick) lines are with (without) the local
contribution of the DM inside the halo. Note that the thin dotted line
is completely superimposed to the thick dotted line, because the local
contribution of decaying DM is always negligible. Top panel: isothermal
halos. Bottom panel: NFW halos.}
\end{figure}
\subsection{Concentration and local contribution}
In this paper we basically considered two different DM profiles:
isothermal and NFW
halos. These two models are comparable in terms of virial radius
and main properties; but the NFW one is much more concentrated. From the comparison between Fig.~\ref{fig:fig5} and
Fig.~\ref{fig:fig5bis} we have already seen that this difference has important
effects on the critical mass: $m_{crit}$ is generally much lower for a
NFW than for an isothermal halo, independently of DM decays and
annihilations. For decaying particles, higher concentrations lead to
marginally stronger effects in delaying the formation of structures, but
the effect is minimal. For annihilations this effect of high
concentrations is much larger (see Figs.~\ref{fig:fig5},
 \ref{fig:fig5bis} and \ref{fig:fig7}).

Another significant characteristic of our model is the estimate of the
local contribution due to DM decays/annihilations occurring inside the
halo (see Section 2.4.2). Pointedly, the local contribution strongly
depends on the DM profile. But what is the importance of the local
contribution? In Fig.~\ref{fig:fig7} we indicate the effect on
$m_{crit}$ of LDM decays (dotted line) and annihilations (dot-dashed),
by including (thin line) or not (thick) the local contribution.  In the
case of DM decays the thin and thick dotted lines appear superimposed,
both in the isothermal (top panel) and NFW (bottom) profile, indicating
that the local contribution is always negligible for decays.

Instead, if we consider the annihilations, the case with (thin line) and
without (thick) the local contribution are very different, especially
for
the NFW profile. If we do not include the local contribution, $m_{crit}$
is very close to the unperturbed value. If we switch on the
local contribution in the isothermal profile, $m_{crit}$ is
substantially higher (a factor $\sim{}2$) than in the unperturbed case,
at least for high virialization redshifts ($z_{vir}>30$). Finally, if we
account for the local contribution in a NFW halo, $m_{crit}$ is always
higher (a factor $\sim{}2-4$) than in the unperturbed case (even at low
virialization redshifts). This is consistent with the fact that the
annihilation rate strongly depends on the local DM density.

\section{Conclusions}
In this paper we derived the effects of DM decays and annihilations on structure formation. We considered only moderately massive DM particles (sterile neutrinos and LDM), as they are expected to give the maximum contribution to heating and reionization (Mapelli et al. 2006). To describe the interaction between the IGM and the decay/annihilation products we followed the recipes recently derived by RMF06.

We accounted not only for the diffuse cosmological contribution to
heating and ionization, but also for the local contribution due to DM
decays and annihilations occurring in the halo itself. The local
contribution results to be dominant in the case of DM annihilations
especially for cuspy DM profiles.

The energy injection from DM decays/annihilations produces both an
enhancement in the abundance of coolants (H$_2$ and HD) and an increase
of gas temperature.  We found that for all the considered DM models
(sterile neutrino decays, LDM decays and annihilations) the critical
halo mass for collapse, $m_{crit}$ is often higher than in the
unperturbed case. 
 This means that DM decays and annihilations
tend to delay the formation of structures. However, the variation of
$m_{crit}$ is minimal. In the most extreme cases, i.e. considering LDM
annihilations (decays) and halos virializing at redshift $z_{vir}>30$
($z_{vir}\sim{}10$), $m_{crit}$ increases of a factor $\sim{}4$
($\sim{}$2). 
In the case of decays, the variations of $m_{crit}$ are almost
independent from the assumed concentration of the DM halo, although
higher concentrations (corresponding to smaller values of $m_{crit}$)
seem to be associated with slightly stronger effects of the DM energy
injection. The dependence on concentration is more evident in the case
of annihilating particles, where higher concentrations lead to
substantially larger effects. This happens because the ``local''
contribution is important.

In summary, the effects of DM decays and/or annihilations on structure
formation are quite small, except in some extreme cases (e.g. very
high concentration for annihilations). However, the energy injection from DM
decays/annihilations has important consequences on the fraction of gas
which is retained inside the halo. This fraction can be substantially
reduced, especially in the smallest halos
($\lesssim{}10^6\,{}M_\odot{}$).

Finally, we point out that our results are quite different from the
conclusions of Biermann \& Kusenko (2006) {and Stasielak et
al. (2006)\footnote{Biermann \& Kusenko (2006) and Stasielak et
al. (2006) do not calculate the critical mass $m_{crit}$. So, it is
quite difficult to make a quantitative comparison between their results
and ours.}, who suggest that sterile neutrino decays can favour the
formation of first objects.  The discrepancy is likely due to our more
complete treatment which includes the hydrodynamics of the collapsing
structures. In fact, our hydro-dynamical treatment allows to describe
the detailed gas density evolution during the collapse, resulting in
markedly different temperature and chemical properties with respect to
those found by a simple one-zone model.



\section*{Acknowledgements}
We thank S. Zaroubi, C. Watson, A. Kusenko, P. Biermann and J.~Stasielak for useful discussions. We also acknowledge the referee for a critical reading of the paper. The three authors
thank both SISSA/ISAS and the Kapteyn Institute for mutual hospitality
during the preparation of this paper. ER gratefully acknowledges support
from the Netherlands Organization for Scientific Research (NWO) under
project number 436016.


\onecolumn
\appendix

\end{document}